# Optically synchronized fiber links with spectrally pure integrated lasers

Grant M. Brodnik[1], Mark W. Harrington[1], John H. Dallyn[2], Debapam Bose[1], Wei Zhang[3,a], Liron Stern[3,b], Paul A. Morton[4], Ryan O. Behunin[2,5], Scott B. Papp[3,6], and Daniel J. Blumenthal[1,*]

[1] Department of Electrical and Computer Engineering, University of California Santa Barbara, Santa Barbara, CA, USA
[2] Department of Applied Physics and Material Sciences, Northern Arizona University, Flagstaff, Arizona, USA
[3] Time and Frequency Division 688, National Institute of Standards and Technology, Boulder, CO, USA
[4] Morton Photonics, West Friendship, MD, USA
[5] Center for Materials Interfaces in Research and Applications (¡MIRA!), Northern Arizona University, Flagstaff, Arizona, USA
[6] Department of Physics, University of Colorado, Boulder, Boulder, CO, USA
[a] Current address: Jet Propulsion Laboratory, Pasadena, CA, USA
[b] Current address: Department of Applied Physics, Hebrew University of Jerusalem, Israel

* Corresponding author (danb@ucsb.edu)

## ABSTRACT

Precision frequency and phase synchronization between distinct fiber interconnected nodes is critical for a wide range of applications, including atomic timekeeping, quantum networking, database synchronization, ultra-high-capacity coherent optical communications and hyper-scale data centers. Today, many of these applications utilize precision, tabletop laser systems, and they would benefit from integration in terms of reduced size, power, cost, and reliability. In this paper we report a record low $3 \times 10^{-4}$ rad$^2$ residual phase error variance for synchronization based on independent, spectrally pure, ultra-high mutual coherence, photonic integrated lasers. This performance is achieved with stimulated Brillouin scattering lasers that are stabilized to independent microcavity references, realizing sources with 30 Hz integral linewidth and a fractional frequency instability $\leq 2 \times 10^{-13}$ at 50 ms. This level of low phase noise and carrier stability enables a new type of optical-frequency-stabilized phase-locked loop (OFS-PLL) that operates with a < 800 kHz loop bandwidth, eliminating traditional power consuming high bandwidth electronics and digital signal processors used to phase lock optical carriers. Additionally, we measure the residual phase error down to a received carrier power of -34 dBm, removing the need to transmit in-band or out-of-band synchronized carriers. These results highlight the promise for a path to spectrally pure, ultra-stable, integrated lasers for network synchronization, precision time distribution protocols, quantum-clock networks, and multiple-Terabit per second coherent DSP-free fiber-optic interconnects.

# INTRODUCTION

The distribution of precision optical frequency and phase references over fiber optic links[1–6] enables applications including high-precision metrology, investigation of variations in fundamental physical constants[7], interferometry for radio astronomy[8,9], optical clock comparison[10–13] and high-capacity coherent communications[14]. To achieve high precision, scientific applications employ ultra-low linewidth, frequency stabilized lasers[15,16] that require sophisticated setups at the tabletop scale and high capacity coherent communications employ power intensive digital signal processors (DSP)[17,18]. Miniaturizing these stable lasers and reducing the power of architectures that employ DSP will enable applications including distributed network synchronization[19], precision time protocols[20], quantum optical clock networks[21], and energy-efficient, high-capacity coherent fiber optic interconnects[22,23]. Therefore, solutions are needed to reduce the cost, size, complexity, and power consumption of ultra-stable, ultra-low linewidth lasers that are able to support these precision fiber links.

Optical frequency and phase synchronization between integrated transmit and receive lasers over coherent fiber links is traditionally accomplished by use of high speed feedback techniques like electronic phase locked loops (EPLLs)[24,25], optical phase locked loops (OPLLs)[26–30], or real time frequency and phase recovery with DSPs[17,18,31,32]. However, as applications drive a need to improve the precision of phase and carrier synchronization, new techniques are needed that scale favorably in terms of complexity, power consumption, and cost to overcome limitations such as the bandwidth of electronics in DSPs used in digital intradyne coherent systems[33,34], and the feedback loop requirements[35] and high frequency analog and digital electronics used in OPLLs and EPLLs. An alternative approach is to leverage the spectral purity of ultra-stable, ultra-low linewidth lasers, traditionally used in high-end precision scientific experiments[1–5,14,15]. These lasers combine feedback techniques such as the Pound-Drever-Hall (PDH)[36] method with environmentally isolated, thermally engineered glass or single crystal high-Q optical reference cavities[15,16,37]. For example, state of the art, lab-scale lasers demonstrating linewidths below 10 mHz with a carrier stability of $4\times10^{-17}$ at timescales between one and a few tens of seconds[38] have been reported. Miniature, chip-scale stabilized lasers, with exceptional noise and stability characteristics, have been realized by locking semiconductor lasers or Brillouin scattering (SBS) lasers to bulk-optic whispering gallery mode resonators[39,40], photonic integrated spiral waveguide delay lines[41], and microcavity Fabry-Perot resonators[22,42] or by locking the transmit laser to an atomic transition[14]. However, to date, there has not been demonstration of a precision phase synchronized fiber link that is based on independent, compact, spectrally pure stabilized lasers, where the lasers' mutual coherence yields ultra-low residual phase error with significantly less electrical bandwidth than traditional approaches.

In this paper we report, to the best of our knowledge, the first demonstration of precision frequency and phase-locked optical sources over a fiber-optic link employing independent chip-scale, frequency stabilized integrated lasers with ultra-high mutual coherence. The fiber-connected lasers are phase synchronized with a low residual phase error variance of $3\times10^{-4}$ rad$^2$ for a homodyne lock, orders of magnitude lower than achieved with OPLLs employing large linewidth integrated lasers[28]. This performance is achieved by combining stabilized, spectrally pure, independent chip-scale lasers with an optical-frequency-stabilized phase-locked loop (OFS-PLL). Synchronization is achieved using only low-bandwidth OFS-PLL lock loops (<800 kHz) in the individual laser stabilization and the receiver phase lock circuits without the need for the high bandwidth electronics of EPLL or OPLL circuits or high-power consumption DSPs. The lasers are a cavity stabilized SBS laser (CS-SBS) design, consisting of a photonic integrated silicon nitride (SiN) SBS laser[43] locked to a silica Fabry-Perot optical microcavity[42]. We measure the optical frequency noise (FN) of the transmit (Tx) and receive (Rx) lasers, as well as the heterodyne beatnote FN spectrum between Tx and Rx lasers, resulting in an integral linewidth of ~30 Hz for each CS-SBS laser and a carrier fractional frequency stability of $2\times10^{-13}$ at 50 ms. For the homodyne phase lock, the low residual phase error is achieved with received optical powers (ROP) as low as -34 dBm, eliminating the need for out of band optical carriers[44]. These results represent a major leap toward deploying chip-scale stabilized photonic integrated lasers for applications requiring precision carrier and phase recovery over fiber links, using low bandwidth control and feedback circuitry that consumes on the order of several mWs when implemented in bipolar complementary metal-oxide-semiconductor (BiCMOS) circuits, without the need for high power, high bandwidth analog electronics or DSPs, opening up a new class of precision applications with chip scale photonics.

# RESULTS

**Precision Fiber Link and Node Architecture:** A precision, optically synchronized link is illustrated in Fig. 1. At each end, a CS-SBS laser is used for transmit and receive signals. The microcavity-stabilized stimulated Brillouin scattering design is a tunable semiconductor pump laser[45] locked to a silicon nitride waveguide SBS laser using a low bandwidth locking PDH optoelectronic loop[43]. The sub-Hz fundamental linewidth SBS laser output is modulated by an acousto-optic modulator (AOM), to generate a tunable wavelength that has similar phase noise and linewidth properties to that of the SBS laser. The AOM frequency is driven by a voltage-controlled oscillator (VCO) and tuned to align and lock the optical carrier to a resonance of the ultra-stable silica reference microcavity in a second low bandwidth PDH lock circuit. This reference cavity lock stabilizes the SBS optical carrier, reducing noise within the loop bandwidth of several hundred kHz and reducing the integral linewidth[42]. The spectrally pure SBS emission is split to create a stabilized Tx and Rx-side tunable optical local oscillator (LO) signal. The transmit carrier is coupled to the fiber optic link and sent to a node with a second independent CS-SBS laser of the same design. At the receiving node, the incoming transmitted signal is optically mixed with the local oscillator CS-SBS laser that is made tunable using an optical single sideband modulator or AOM. The photo-mixed signal serves as the error signal for a homodyne optical phase lock[27]. The error signal is filtered by a low bandwidth <800 kHz proportional-integral$^2$-derivative (PI$^2$D) loop filter with integral poles at 50 kHz and 100 kHz and fed back to the LO tuning control to close the phase lock, realizing the OFS-PLL.

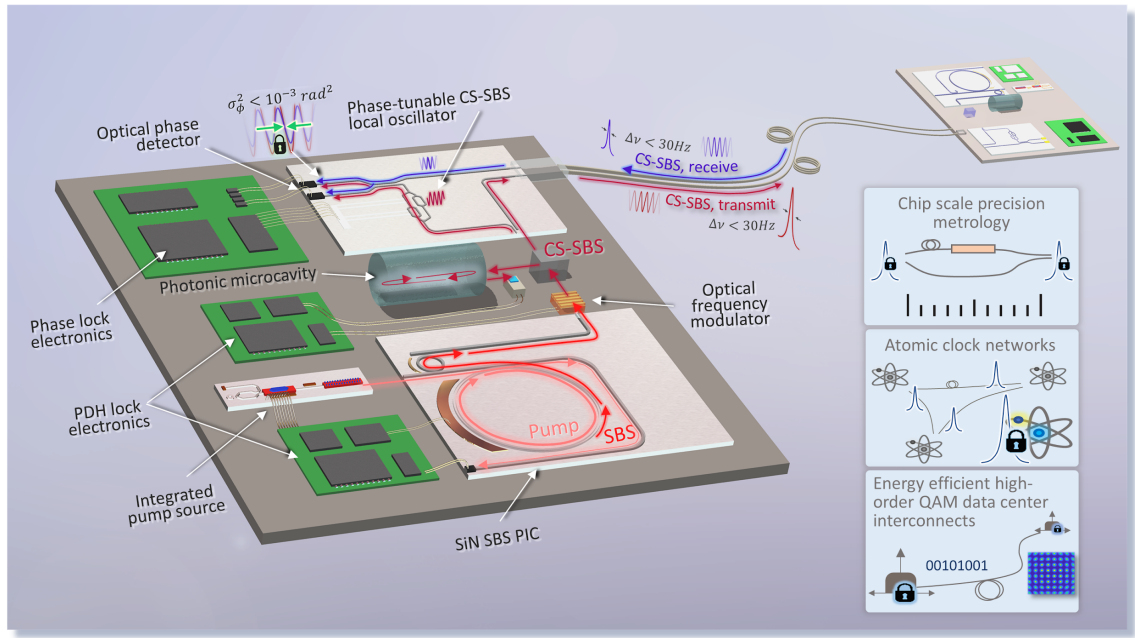

**Fig. 1 | Optically synchronized precision fiber link.** In a two-laser system, independent cavity-stabilized stimulated Brillouin scattering (CS-SBS) lasers provide ultra-low frequency noise and narrow linewidth, stabilized carriers to enable high spectral purity carrier phase and frequency distribution over fiber. SBS lasing is performed in an ultra-low optical loss silicon nitride (SiN) photonic integrated circuit (PIC)[43] for phase noise reduction at all offset frequencies and a corresponding decrease in fundamental linewidth to <1 Hz. Carrier stabilization is implemented with low bandwidth sub-MHz Pound-Drever-Hall (PDH) feedback loops. CS-SBS lasers source ~30 Hz integral linewidth, stable carriers for the optical frequency stabilized phase-lock loop (OFS-PLL) chip-scale architecture, for field deployed applications in precision metrology and high-capacity coherent quadrature amplitude modulation (QAM) communications.

The CS-SBS laser is shown in further detail in Fig. 2. Linewidth narrowing occurs in two stages, where the second stage also performs carrier stabilization. In the first stage, the SBS laser reduces the fundamental linewidth of an external grating semiconductor pump laser from ~100 Hz to just over 1 Hz based on the Brillouin lasing dynamics[46]. The pump laser is aligned to a SiN optical cavity resonance, and a low-bandwidth PDH lock tracks the resonance for efficient transfer of pump power to first order SBS lasing. The resulting SBS emission has low frequency noise of around 10 $Hz^2Hz^{-1}$ at offsets near ~100 kHz, dropping to approximately ~1 $Hz^2Hz^{-1}$ at offsets greater than 1 MHz from the carrier.

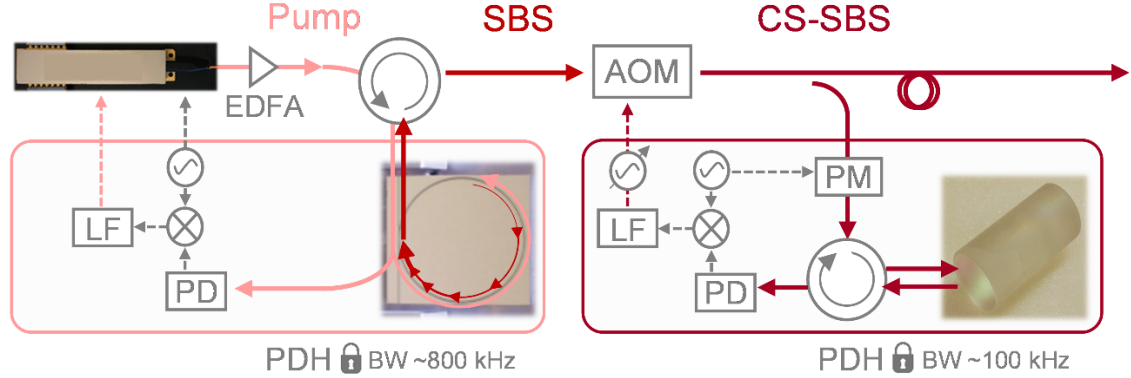

**Fig. 2 | Cavity-stabilized SBS laser (CS-SBS). a**, An ultra-narrow linewidth external fiber cavity diode laser (ECDL)[45] is optically amplified using an erbium-doped fiber amplifier (EDFA) to serve as the stimulated Brillouin scattering (SBS) laser pump. The SBS laser[43] is stabilized to the chip-scale microcavity[42] using an acousto-optic modulator (AOM) operated as an optical frequency shifter with a variable frequency drive signal from a voltage-controlled oscillator. Pound-Drever-Hall (PDH) loops incorporate phase modulators (PM) driven by a voltage-controlled oscillator and photodiodes (PD), electronic mixers, and low bandwidth (BW) loop filters (LF).

For the second stage of linewidth reduction and carrier stabilization, we utilize a >1 Billion-Q Fabry-Perot microcavity housed in a vibration and temperature isolation enclosure[42]. This 1-inch silica microcavity significantly reduces close-to-carrier (CTC)[47] noise sources (below ~100 Hz offset from carrier), including environmental and technical noise[41] as well as conversion of erbium-doped fiber amplifier (EDFA)-induced pump relative intensity noise (RIN) to photothermal noise[41,47,48] inside the SBS cavity. The thermal mass of this microcavity is large, with a time constant for thermal conduction to the environment on the order of 5 minutes, and therefore it does not respond directly to external macroscopic control on the timescale of these frequency noise sources (e.g. changing the temperature of the reference cavity). The control loop must be able to adjust the optical frequency in the CTC to mid-frequency range (DC to 50 kHz for the transmit, Tx, and DC to 100 kHz for the receive, Rx) as well as satisfy control loop phase stability conditions. We use an AOM to create a tunable optical carrier that is locked to the reference cavity and responds in this frequency range. Other techniques are possible including controlling the resonator optical mode photothermally by small adjustments in the lock signal optical power inside the reference cavity[40,42], and are subject of future investigation in this architecture. The PDH feedback signal controls the frequency of a VCO that in turn drives the AOM. When the VCO control servo signal approaches the maximum tuning range of the AOM, the temperature of the SiN SBS laser is adjusted to maintain the VCO and AOM within the lock range. The result is a very low frequency PDH loop that is able to reduce SBS laser frequency noise by more than 3 orders of magnitude, from $10^4$ to $10^1$ $Hz^2Hz^{-1}$. These PDH loops, with only several hundred kHz bandwidth, employ $PI^2D$ feedback filters that if implemented in BiCMOS will only consume on the order of several mWs[49], making this an energy efficient technique for frequency and phase lock at the 100s of THz optical carrier frequencies.

**Frequency Noise, Linewidth and Stability:** We characterize the laser frequency noise and carrier drift at each node and the heterodyne beatnote at the receiver in order to measure the mutual coherence, shown in Fig. 3. The following parameters are extracted from the frequency noise measurements (as described in the Methods Section): The fundamental linewidth (FLW), integral linewidth (ILW), and Allan deviation (ADEV) of the fractional frequency instability (FFI). Optical frequency noise is measured using a radiofrequency (RF)-calibrated fiber-based asymmetric Mach-Zehnder interferometer (aMZI) as an optical frequency discriminator (OFD)[50]. The OFD measurements are valid down to 10 kHz offsets at which fiber noise in the aMZI dominates. Below 10 kHz offset, we use an electronic frequency discriminator (EFD) to measure frequency noise via heterodyne between the transmit and receive pump, SBS, and CS-SBS lasers. The EFD method[51], described in further detail in the Methods section, employs a frequency discriminator to convert frequency fluctuations of the intermediate frequency (IF) signal to voltage fluctuations. The output is sampled by an analog-to-digital converter (ADC) and processed offline for frequency noise and ADEV of fractional frequency instability. This heterodyne measurement between two optical carriers captures the combined noise dynamics of each independent optical source. For the OFD traces of the pump, SBS, and CS-SBS lasers, we sum the frequency noise power spectral densities and stitch into the EFD measurements to represent heterodyne laser noise for the full frequency span of 1 Hz to 10 MHz. The frequency noise measurements are presented in Fig. 3b. Further details of the OFD and EFD measurement systems are provided in Supplementary Information and shown in Supplementary Fig. 1.

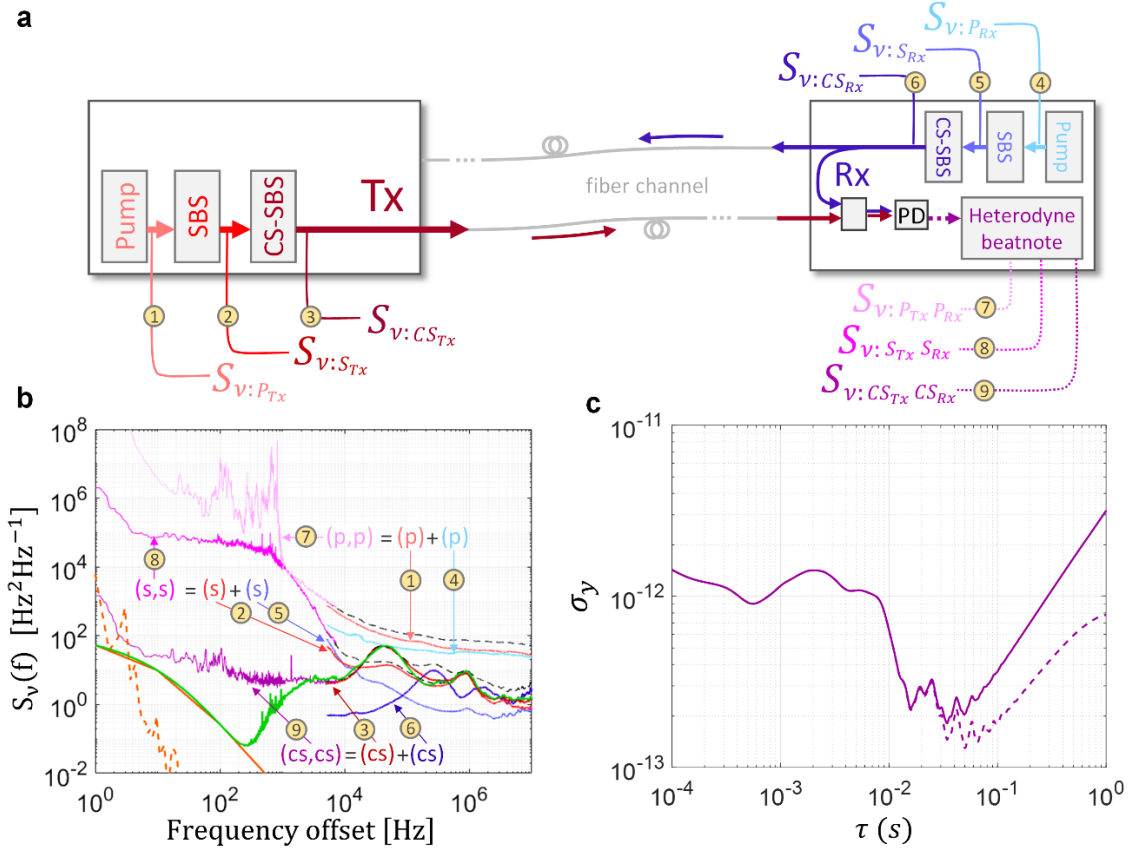

**Fig. 3 | Laser carrier and heterodyne beatnote linewidth and stability measurements. a,** Two independent cavity-stabilized stimulated Brillouin scattering (CS-SBS) lasers (transmit, Tx, and receive, Rx) are connected by optical fiber and photomixed at the receiver on a photodiode (PD) yielding a heterodyne beatnote signal. Frequency noise (FN), $S_\nu(f)$, of the Tx and Rx pump, Tx and Rx SBS, and Tx and Rx CS-SBS optical carriers and receiver-side heterodyne beatnotes between Tx-Rx pump, Tx-Rx SBS, and Tx-Rx CS-SBS are labeled 1-9 at their measurement locations and correspond to traces in (b). **b,** FN of the independent optical carriers (pump (p), SBS (s), and CS-SBS (cs) = light, medium, and dark red (blue) for Tx (Rx), respectively) measured with an optical frequency discriminator (OFD), and of the beatnote between Tx-Rx pump (p,p), Tx-Rx SBS (s,s), and Tx-Rx CS-SBS (cs,cs) lasers (light, medium, and dark purple respectively) measured with an electrical frequency discriminator (EFD). Black dashed lines sum the FN of the independent optical carriers for each configuration. Microcavity thermal noise limit (orange, solid) and photothermal (orange, dashed) are shown. FN of a CS-SBS laser limited only by cavity thermal noise is calculated (green). **c,** Overlapping Allan deviation of the fractional frequency instability of the Tx-Rx CS-SBS heterodyne beatnote referenced to a 193 THz carrier frequency, with/without (solid/dotted) linear drift of 267 Hz/s.

From the frequency noise spectra, we extract the white noise floor, $h_0$ [Hz$^2$Hz$^{-1}$], at high offset frequencies to calculate FLW as $\pi h_0$ [Hz]. The FLWs of each the Tx and Rx pump lasers (light red (p) and light blue (p) in Fig. 3b) are measured to be ~60 Hz. After the SBS lasing stage, the SBS lasers (red (s), blue (s) in Fig. 3b) are measured to have fundamental linewidths reduced to several Hz. Following the microcavity stabilization of the SBS lasers, the FLWs of each CS-SBS laser (dark red (cs) and dark blue (cs) in Fig. 3b) as measured by the high frequency offset FN slightly increase to ~10 Hz. We observe servo bumps as the phase approaches 180° and positive feedback near the PI$^2$D loop bandwidths in the microcavity lock loops at ~50 kHz and ~100 kHz for the Tx and Rx CS-SBS lasers, respectively. The increase in measured fundamental linewidth is a result of this gain peaking and associated increase in FN in the region we extract the white noise floor to calculate the FLW. At high frequency offsets, the FN is expected to fall back to several Hz$^2$Hz$^{-1}$ as achieved by the SBS lasing stage.

In order to understand the fundamental limits of microcavity stabilization on CS-SBS frequency noise, we simulate the SBS stabilization stage using an ideal PDH lock by mapping experimental SBS frequency noise through a PI$^2$D feedback loop with poles matched to those implemented in the experimental setup. In this scenario, the laser noise closely tracks the microcavity frequency reference and the simulated CS-SBS (green curve in Fig. 3b) in-loop frequency noise is limited only by cavity fundamental thermal noise. Discussion of the simulation is given in more detail in Supplementary Information.

We employ two methods to calculate integral linewidths (ILW) of the heterodyne beatnote signals: the $\beta$-separation method[52] and an integration of phase noise from high offset frequencies down to the frequency offset at which the integrated phase noise equals $1/\pi$ rad$^2$ [6,53]. Further details of the ILW calculations are provided in Methods and Supplementary Information and shown in Supplementary Fig. 3. The integral linewidths of the heterodyne beatnote between Tx-Rx pump lasers are 133 kHz and 2.97 kHz calculated using the $\beta$-separation and $1/\pi$ methods, respectively. After SBS lasing, the heterodyne beatnote (Tx-Rx SBS) ILW drops to 15.1 kHz ($\beta$) and 1.97 kHz ($1/\pi$) and is dominated by photothermal noise. The microcavity stabilization stage reduces CTC and intermediate offset noise of the Tx and Rx carriers up to the PDH loop bandwidths of 50 kHz and 100 kHz, respectively, and we measure a decrease in ILW of the Tx-Rx CS-SBS heterodyne down to 104 Hz ($\beta$) and 43 Hz ($1/\pi$). Assuming equal contribution by each CS-SBS laser to the heterodyne 43 Hz ($1/\pi$) integral linewidth, each independent optical integral linewidth is ~30 Hz. Based on the OFD measurements in Fig. 3b, we note that the FN of the Tx CS-SBS (dark red, 'cs') laser contributes more to the heterodyne linewidth based on its higher CTC noise than the independent Rx CS-SBS (dark blue, 'cs'). FLWs and ILWs are summarized in Supplementary Table 1. The ADEV FFI of the Tx-Rx CS-SBS heterodyne in Fig. 3c reaches a minimum value of 2x10$^{-13}$ at a timescale of 50 ms and to just over 1x10$^{-13}$ when a linear drift of 267 Hz is removed. Lineshape evolution as a function of observation time and corresponding ADEV FFI are described in more detail in Supplementary Information and in Supplementary Fig. 2.

**Precision, Optically Synchronized Fiber Link:** The combined linewidth and open loop heterodyne noise characteristics between two optical sources quantify the phase noise and frequency drift that must be compensated when closing the phase-locked-loop. Leveraging CS-SBS linewidth narrowing and stabilization, the OFS-PLL relaxes these phase lock demands and improves phase lock performance in comparison to standard phase locks that align the optical phases and suppress phase noise only within their loop bandwidths and with reduced gain.

The OFS-PLL architecture is shown in Fig. 4a below where we synchronize the optical phase between two independent CS-SBS Tx and Rx lasers over optical fiber. The Tx and Rx (tunable local oscillator, LO) signals are photomixed on a balanced photodetector to generate a homodyne phase error signal[27]. The PI$^2$D filtered phase error signal is used to control a 6 GHz VCO that drives a phase modulator with an optically filtered upper sideband to achieve optical frequency tuning required for the homodyne phase lock. While discrete components were employed here for frequency tunability in this proof of concept demonstration, integrated single sideband modulators have been demonstrated in silicon photonic (SiPh) platforms[54] with low power consumption and high carrier suppression for fully integrated low power operation of the OFS-PLL.

We show the phase error signal between mutually coherent CS-SBS lasers in Fig. 4b for open-loop OFS-PLL, closed-loop OFS-PLL, and the lock-in transient between these regimes as measured on a digital oscilloscope (DSO). In open-loop operation, the phase error signal is simply the heterodyne beatnote of the two CS-SBS lasers, oscillating at the difference frequency of the Tx and Rx optical sources and with the measured FN characteristics shown previously in Fig. 3b. The open-loop to closed-loop transient shows a zero-fringe slip transition to the optical phase locked state representing successful lock-in acquisition[55]. We estimate a phase lock-in range of 700 kHz with repeated lock acquisition attempts and increasing initial frequency differences between optical Tx and Rx CS-SBS lasers until the phase lock cannot engage. This initial OFS-PLL design does not include a coarse frequency pull-in loop to expand the region of phase lock engagement. During closed-loop operation, the root-mean-square (RMS) voltage error is used to calculate the residual phase error variance[56] of $3\times10^{-4}$ rad$^2$ with received optical powers as low as -34 dBm. To demonstrate long-term and robust OFS-PLL operation, we measure residual phase error over a span of 20 minutes for varied received optical powers between -14 dBm and -44 dBm in Fig. 4c. At low ROP of -44 dBm, the residual phase error increases to $\sim 8\times10^{-3}$ rad$^2$. In Fig. 4d, we show a sample 16-symbol quadrature-amplitude modulation (16-QAM) constellation to illustrate a large phase noise impairment on recovered symbols in a coherent optical communications link. The experimentally measured OFS-PLL low phase error variance of $3\times10^{-4}$ rad$^2$ is sufficient for high order QAM (e.g. 512-QAM) in a phase noise limited link[57].

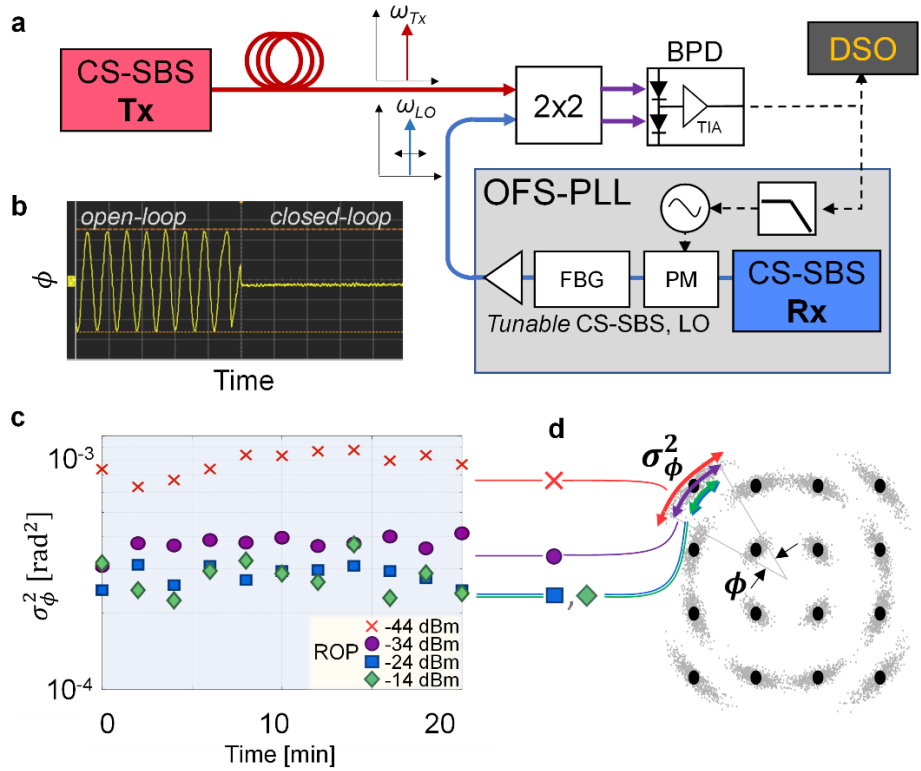

**Fig. 4 | Precision link operation and performance. a**, A microcavity-stabilized stimulated Brillouin scattering (CS-SBS) laser transmits (Tx) a linewidth narrowed, stable optical frequency reference over fiber. A CS-SBS local oscillator (LO) is made tunable by optically filtering (via fiber Bragg grating, FBG) a phase modulated (PM) sideband driven by a voltage-controlled oscillator with error signal derived from a balanced photodiode (BPD) with built in transimpedance amplifier (TIA) as a homodyne phase detector[27]. **b**, Digital storage oscilloscope (DSO) time trace of optical-frequency-stabilized phase-locked loop (OFS-PLL) phase error signal demonstrating open-loop, lock acquisition transient, and closed-loop operation. **c**, OFS-PLL residual phase error variance, $\sigma_\phi^2$, over multiple 20 min time scales at CS-SBS received optical power (ROP) down to -44 dBm at the phase detector. **d**, in an example 16-QAM IQ plane, $\sigma_\phi^2$ describes noise normally distributed angularly about the origin for each recovered symbol (exaggerated for illustrative purposes). The achieved OFS-PLL residual phase error variance measured in (c) is sufficient for high data capacity, high order QAM transmission (e.g. 512-QAM)[57].

# DISCUSSION

We demonstrate that the ultra-high mutual coherence between photonic integrated, spectrally pure lasers can be used to achieve high-precision, ultra-low residual phase error variance over optical fiber by employing techniques traditionally relegated to high-end scientific experiments. These results show promise to bring the performance of lasers that typically occupy table-top environments to a wide range of new fiber-optic distributed applications. The ability to provide this level of phase precision synchronization on a 193 THz optical carrier using low bandwidth electronics that are typical of RF links opens a new domain for frequency and wavelength scalable, high-coherence architectures with low energy electronics. We achieve these results by combining SBS dynamics with the narrow resonance and stability of a >1 Billion Q reference microcavity to realize high mutual coherence between independent optical sources, yielding a 43 Hz heterodyne beatnote integral linewidth and fractional frequency instability better than $2 \times 10^{-13}$ at a 50 ms timescale. We also demonstrate that the phase lock for these mutually coherent sources can be achieved using a new type of optical-frequency-stabilized phase-locked loop (OFS-PLL) that achieves a residual phase error variance of $3 \times 10^{-4}$ rad$^2$ with loop bandwidth < 800 kHz and low received optical power of -34 dBm. The individual photonic components of the OFS-PLL are readily integrated in silicon nitride (SiN) and heterogeneous silicon photonic (SiPh) platforms[54,58,59] with mW-tier BiCMOS electrical integrated circuits. Future work includes OFS-PLL implementations with heterogeneously integrated lasers as the SBS pump source[60], waveguide integration of the reference cavities[41,61,62], and co-location of the control electronics[49] to provide a fully integrated, low-power solution. Additional modifications to the OFS-PLL architecture include adding a frequency pull-in stage to extend the capture range of the optical phase lock and operating with both polarizations to enable polarization diverse sensing and dual polarization optical communications. The photonic integrated CS-SBS laser source can be used to pump integrated optical nonlinear microresonators to realize high spectral purity optical frequency comb sources at the chip scale[63] for metrology and generation of wave-division-multiplexed (WDM) carriers for DSP-free coherent communications. These results open new applications that require ultra-high spectral purity laser sources and precision optical phase in compact, low power implementations, such as distributed atomic clocks for timing and metrology and energy efficient DSP-free Tbps coherent optical links in datacenter interconnects[22,23].

# METHODS

**Frequency noise of optical carriers:** We directly measure the power spectral density (PSD) of frequency fluctuations (frequency noise), $S_\nu(f)$, of independent optical sources using a calibrated, optical fiber-based asymmetric Mach-Zehnder interferometer (aMZI) optical frequency discriminator. Optical frequency fluctuations of the laser under test are converted into analog voltage fluctuations that are sampled by an analog-to-digital converter (ADC) and processed offline for spectral characterization. The fiber-based aMZI consists of a 200 m fiber delay in the asymmetric path and a corresponding measured free spectral range (FSR) of 1.026 MHz. The ADC sampling is performed using a real-time scope triggered when the optical source is at the quadrature region of the aMZI transfer function. With known peak-to-peak voltage fluctuations, $V_{pp}$, and aMZI optical time delay, $\tau_d$, the voltage PSD, $S_V(f)$ in V$^2$Hz$^{-1}$ is converted to frequency noise, $S_\nu(f)$, in units of Hz$^2$Hz$^{-1}$ as a function of carrier offset frequency, $f$, by (1) below[39]:

$$S_\nu(f) = S_V(f) \left[ \frac{f}{\sin(\pi f \tau_D) V_{pp}} \right]^2 \qquad (1)$$

This aMZI transfer function conversion between PSDs results in artificial spurs at integer multiples of the aMZI FSR due to the $\sin(\cdot)$ term in the denominator, which are removed with a median filter in the data processing step. Additional details of the OFD measurements are provided in Supplementary Information and schematically in Supplementary Fig. 1.

**Frequency noise of optical heterodyne beatnote:** Optical frequency discriminator FN measurements are dominated by fiber and vibration noise in the optical delay path at carrier offset frequencies below approximately 10 kHz, requiring an additional method to accurately report $S_\nu(f)$ close-to-carrier. We do this by optically mixing two lasers and measuring the frequency noise of the intermediate frequency (IF) signal with a Miteq FMDM electrical frequency discriminator[51]. The voltage output scaled by the discriminator constant is the frequency error of the IF signal. The output is filtered with a 5th order pi-filter with bandwidth of 200 kHz to prevent aliasing of high frequency noise into the region of interest (1 Hz to 10 kHz) and then sampled by a high-resolution ADC. Additional details of the EFD measurements are provided in Supplementary Information and schematically in Supplementary Fig. 1. The time domain samples are processed offline for spectral measurements and yield the reported frequency noise and ADEV plots given in Fig. 3b,c. Additional details of the ADEV measurements are provided in Supplementary Information and in Supplementary Fig. 2.

**Fundamental and integral linewidths:** We use the measured frequency noise, $S_\nu(f)$, to calculate the fundamental linewidth of each independent optical source. The fundamental linewidth, also often referred to as the instantaneous or quantum-limited linewidth, gives rise to a laser's Lorentzian lineshape for a white-noise-only laser and can therefore be determined by extracting the white noise floor in the FN spectra far-from-carrier, where environmental noise is no longer dominant. The fundamental linewidth in Hz is then given as

$$\delta_\nu = \pi h_0 \quad (2)$$

where $h_0$ is the white noise floor of $S_\nu(f)$ in units of Hz$^2$Hz$^{-1}$.

Integral linewidths of the optical sources are calculated using two approaches, the $\beta$-separation method[52] and an integration of phase noise method[6,53]. The $\beta$-separation method is a modulation index approach to integral linewidth calculation, accounting for the frequency noise components that contribute to lineshape broadening versus modulation that occurs far-from-carrier that does not increase integral linewidth[52]. The second method[6,53] integrates phase noise from a carrier frequency offset of infinity (or far-from-carrier, such that the maximum frequency is ≫ integral linewidth) to the frequency offset at which the integral equals $1/\pi$ radians$^2$. For a white-noise-only laser, this integration yields the offset frequency corresponding to the full-width-half-maximum (FWHM) of the laser lineshape. Additional details of integral linewidth calculations are provided in Supplementary Information and schematically in Supplementary Fig. 3. Linewidth results are summarized in Supplementary Table 1.

**Optical phase lock of cavity stabilized SBS lasers:** One CS-SBS laser is transmitted over fiber and optically mixed with a tunable CS-SBS receive laser. We achieve tunability of the Rx CS-SBS laser using an electro-optic modulator (EOM) driven by a voltage-controlled oscillator to generate phase modulated optical sidebands. Using a fiber Bragg grating optical filter, the upper sideband (USB) is filtered out and the carrier and lower sideband are suppressed. By tuning the frequency of the VCO, the filtered CS-SBS USB is made tunable. The tunable CS-SBS USB is homodyne photomixed with the transmitted CS-SBS carrier signal on a balanced photodetector to generate a phase error signal[27] between the transmit and receive lasers. The phase error signal is passed through a PI$^2$D filter with integrator poles at 50 kHz and 100 kHz and fed back to the VCO control port to close the phase locked loop. The lock-in range[55] is estimated by detuning the initial frequency offset between the two lasers and attempting to engage closed-loop phase-locked operation with zero cycle slips. It is measured to be ~700 kHz, near the estimated servo loop bandwidth.

**Residual phase error of closed loop OFS-PLL:** We quantify phase lock performance by measuring the phase error signal during closed loop operation and we report the phase error variance[56], $\sigma_\phi^2$ rad$^2$. We measure the peak-peak voltage, $V_{pp}$, of the open-loop homodyne signal (beatnote between the two CS-SBS lasers near zero frequency detuning). During closed loop operation, the phase error signal RMS voltage, $V_{RMS}$ divided by $V_{pp}$ and scaled by $\pi$ (a half-cycle of the IF signal corresponds to $V_{pp}$ and $\pi$ radians phase error) is the phase error RMS in radians, or phase error standard deviation since the error is centered around zero. The square of this is the phase error variance, $\sigma_\phi^2$, in rad$^2$.

# DATA AVAILABILITY

The data that support the plots within this paper and other findings of this study are available from the corresponding author on reasonable request.

# ACKNOWLEDGEMENTS

This material is supported under ***OPEN 2018*** Advanced Research Projects Agency-Energy (ARPA-E), U.S. Department of Energy, under Award Number DE-AR0001042.

# DISCLAIMER

The views and conclusions contained in this document are those of the authors and should not be interpreted as representing official policies of ARPA-E or the U.S. Government. The authors declare no competing financial interests.

# CONTRIBUTIONS

G. M. B, D. J. B., and S. B. P. prepared the manuscript. D. J. B. conceived the optical-frequency-stabilized phase-locked-loop (OFS-PLL) approach. D. J. B., G. M. B., M. W. H., and S. B. P conceived the implemented OFS-PLL architecture. P. A. M. contributed the narrow linewidth integrated optical pump sources. D. B. fabricated the SiN integrated high-Q resonator stimulated Brillouin scattering (SBS) laser chip. W. Z., L. S., and S. B. P. contributed the optical microcavity used in the SBS laser stabilization stage. G. M. B. and M. W. H. performed the experiments, including SBS generation, SBS stabilization to microcavities, and optical phase locking along with the associated noise and performance characterizations. J. H. D. and R. O. B. contributed simulations of laser noise performance and identification of noise contributions from SBS and cavity stabilization physics theory. All authors contributed to analyzing the simulated and experimental results. D. J. B. and S. B. P., supervised and led the scientific collaboration.

# Optically synchronized fiber links with spectrally pure integrated lasers: Supplementary Information

Grant M. Brodnik[1], Mark W. Harrington[1], John H. Dallyn[2], Debapam Bose[1], Wei Zhang[3,a], Liron Stern[3,b], Paul A. Morton[4], Ryan O. Behunin[2,5], Scott B. Papp[3,6], and Daniel J. Blumenthal[1,*]

[1] Department of Electrical and Computer Engineering, University of California Santa Barbara, Santa Barbara, CA, USA
[2] Department of Applied Physics and Material Sciences, Northern Arizona University, Flagstaff, Arizona, USA
[3] Time and Frequency Division 688, National Institute of Standards and Technology, Boulder, CO, USA
[4] Morton Photonics, West Friendship, MD, USA
[5] Center for Materials Interfaces in Research and Applications (¡MIRA!), Northern Arizona University, Flagstaff, Arizona, USA
[6] Department of Physics, University of Colorado, Boulder, Boulder, CO, USA
[a] Current address: Jet Propulsion Laboratory, Pasadena, CA, USA
[b] Current address: Department of Applied Physics, Hebrew University of Jerusalem, Israel

**\*** Corresponding author (danb@ucsb.edu)

## SUPPLEMENTARY INFORMATION

**Fundamental limits of the cavity stabilized stimulated Brillouin scattering laser:** To analytically calculate the frequency noise of the microcavity-stabilized stimulated Brillouin scattering (CS-SBS) laser, we modify the free running phase dynamics first-order SBS laser given in ref.[1]. Accounting for PI$^2$D (proportional, double-integral, derivative) feedback modifies the laser phase dynamics (e.g., see ref.[2]). We model the impact of Pound-Drever-Hall[3] (PDH) stabilization by adding time-delayed (accounting for feedback bandwidth) proportional, double integral, and derivative gain to the free-running phase dynamics where the error signal is defined by $e(t) = \dot{\varphi}(t) - \dot{\varphi}_{ref}(t)$. Here $\dot{\varphi}(t)$ (overdot denotes differentiation with respect to time) is the angular frequency of the CS-SBS and $\dot{\varphi}_{ref}(t)$ is the angular frequency of the reference cavity, both evaluated at time $t$. On resonance, the error signal vanishes and the controller does not apply any changes to the system. Assuming thermo-refractive dominated noise of the reference cavity, we solve for the locked laser phase dynamics to obtain the frequency noise power spectral density (PSD) of the CS-SBS laser. This expression accounts for the impact of the feedback loop as well as the noise sources of the reference cavity. These calculations incorporate experimental data in Fig. 3b (Main), as well as the calculated frequency noise of the reference cavity due to thermo-refractive noise[4]. The results of the calculation can be seen in Fig. 3b (Main) as the green trace. Additionally, a calculation of the photothermal noise[4] was plotted separately (orange dashed line) in Fig. 3b (Main) to account for the behavior of the system at low frequencies.

**Discriminator-based frequency noise measurements:** Frequency noise (FN) power spectral densities, $S_\nu(f)$ in units of Hz$^2$Hz$^{-1}$, of the optical sources was measured using frequency discriminators for carrier offset frequencies ranging from 1 Hz to 10 MHz. At high frequency offsets of 10 MHz down to 10 kHz, FN was measured using an optical frequency discriminator (OFD) based on a fiber optic asymmetric Mach-Zehnder interferometer (aMZI), shown in Supplementary Fig. 1a. The aMZI incorporates a 3 dB optical splitter to form two independent optical paths, one consisting of a 200m optical fiber delay line and polarization controller and the other with a voltage-controlled fiber phase shifter. This 200m delay line design results in a free spectral range (FSR) of 1.026 MHz. OFD measurements were made directly with optical carriers as input, with the fiber phase shifter providing slow adjustments to the relative phase of the optical paths to achieve quadrature biasing of the aMZI. An output 2-by-2 combiner photomixes the optical signal on a Thorlabs PDB450C balanced photodetector (BPD). Optical frequency fluctuations of the laser under test are converted into analog voltage fluctuations that are sampled by an analog-to-digital converter (ADC) and processed offline for spectral characterization. The ADC sampling is performed using a DSAX93204A real-time scope triggered when the optical source is at the quadrature region of the aMZI transfer function. With known peak-to-peak voltage fluctuations, $V_{pp}$, and aMZI optical time delay, $\tau_d$, the voltage PSD, $S_V(f)$ in V$^2$Hz$^{-1}$ is converted to frequency noise, $S_\nu(f)$ in units of Hz$^2$Hz$^{-1}$ as a function of carrier offset frequency, $f$, by (1):

$$S_v(f) = S_V(f)\left[\frac{f}{\sin(\pi f \tau_D)\,V_{pp}}\right]^2 \tag{1}$$

Conversion between PSD domains results in artificial spurs at integer multiples of the aMZI FSR due to the $\sin(\cdot)$ term in the denominator, which are removed with a median filter in the data processing step.

Close-to-carrier (CTC) and intermediate offset frequency noise was measured for the heterodyne beatnote of two independent Tx and Rx pump, SBS, and CS-SBS lasers. We do this by photomixing two independent lasers on a Thorlabs PDB470C and measuring the frequency noise of the intermediate frequency (IF) signal with a Miteq FMDM-160/35-15 electrical frequency discriminator[5], as shown in Supplementary Fig. 1b. The discriminator constant, $D$ [Hz·V$^{-1}$], is measured by step-sweeping a radiofrequency (RF) synthesizer across the EFD to trace out its transfer function and fitting a line to the linear region $\sim 160 \pm 10$ MHz. With the IF signal in the linear region near the center frequency of 160 MHz, the voltage output scaled by the discriminator constant is the frequency error of the IF signal. The output is filtered with a 5th order pi-filter with 3 dB bandwidth of 200 kHz to prevent aliasing of high frequency noise into the region of interest (1 Hz to 10 kHz) and then sampled by a high-resolution 1.25MSa National Instruments USB-6259 ADC. The time domain samples are processed offline for spectral measurements and yield the reported frequency noise (Fig. 3b, Main) and Allan deviation (ADEV) (Fig. 3c, Main) of the fractional frequency instability (FFI).

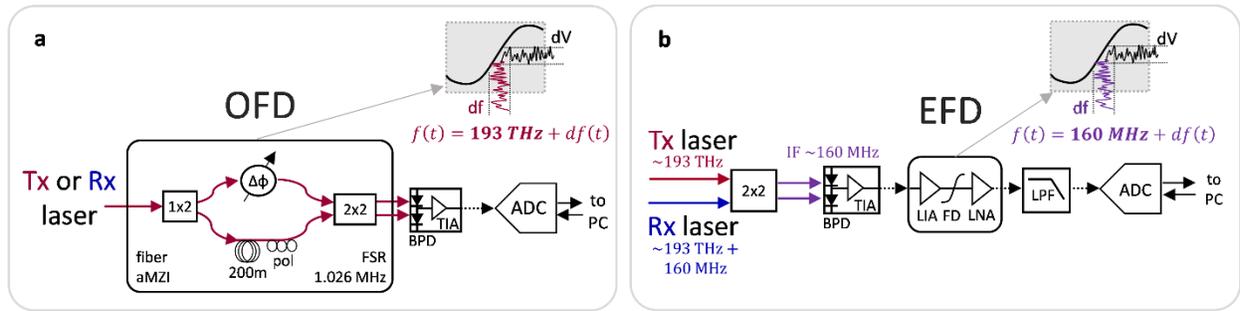

**Supplementary Fig. 1 | Optical/Electrical frequency discriminators. a,** An asymmetric Mach-Zehnder interferometer (aMZI) forms an optical frequency discriminator (OFD)[6] to measure frequency noise of transmit (Tx) and receive (Rx) optical carriers. The aMZI has a 200m fiber delay and corresponding 1.026 MHz free spectral range (FSR). A transimpedance amplified (TIA) balanced photodiode (BPD) converts the optical signal to an electrical signal that is sampled by an analog-to-digital converter (ADC) and processed offline by personal computer (PC). **b,** Tx and Rx lasers with a carrier frequency difference of ~160 MHz are photomixed on a BPD with TIA, yielding a beatnote at intermediate frequency (IF) of ~160 MHz. The IF signal is input to an electrical frequency discriminator (EFD)[5] module consisting of a limiting amplifier (LIA) for amplitude noise insensitivity, a Foster-Seeley circuit frequency discriminator (FD) and low noise amplifier (LNA). The output voltage signal is low pass filtered (LPF) to prevent aliasing, and then sampled by ADC and processed offline.

**Stability of optical heterodyne beatnote:** We photomix two CS-SBS lasers and use the EFD system described in Supplementary Fig. 1a to measure the ADEV FFI of the heterodyne beatnote, shown in Fig. 3c (Main) and in Supplementary Fig. 2. Being a measurement of a two-laser heterodyne, the measured beatnote has noise characteristics of both lasers. In the case of two independent CS-SBS lasers, we assume uncorrelated noise and $1/\sqrt{2}$ contribution of each laser's linewidth to the heterodyne linewidth. In Supplementary Fig. 2a, we illustrate that laser lineshape is a function of observation time and that lineshape characteristics manifest in the ADEV FFI trace in Supplementary Fig. 2b. At short timescales (<1 μs, or correspondingly, high-frequency fluctuations >1 MHz) the laser's white noise floor determines the fundamental linewidth (or FLW, shown as the Lorentzian lineshape in orange and labeled $\boldsymbol{\Delta\nu_{FLW}}$, and the orange region of the ADEV plot). In this region, SBS linewidth narrowing dynamics determine the CS-SBS laser noise and fundamental linewidth. At intermediate timescales (100 μs to ~100 ms, or frequency fluctuations 0.1 Hz to 10 kHz), the laser lineshape is broadened from the fundamental linewidth to an effective, or integral linewidth (ILW) that is a function of observation time. We show this as the light blue lineshape, labeled $\boldsymbol{FWHM(t)}$, and light blue region of the ADEV. In this region, CS-SBS noise is determined by environmental effects and noise limitations of components in the PDH lock stabilizing the SBS carrier to the photonic microcavities. At long timescales (>100ms, or frequency fluctuations <0.1 Hz), the locked SBS carriers faithfully follow the microcavity references, and drift of the microcavities dominates the CS-SBS laser stability.

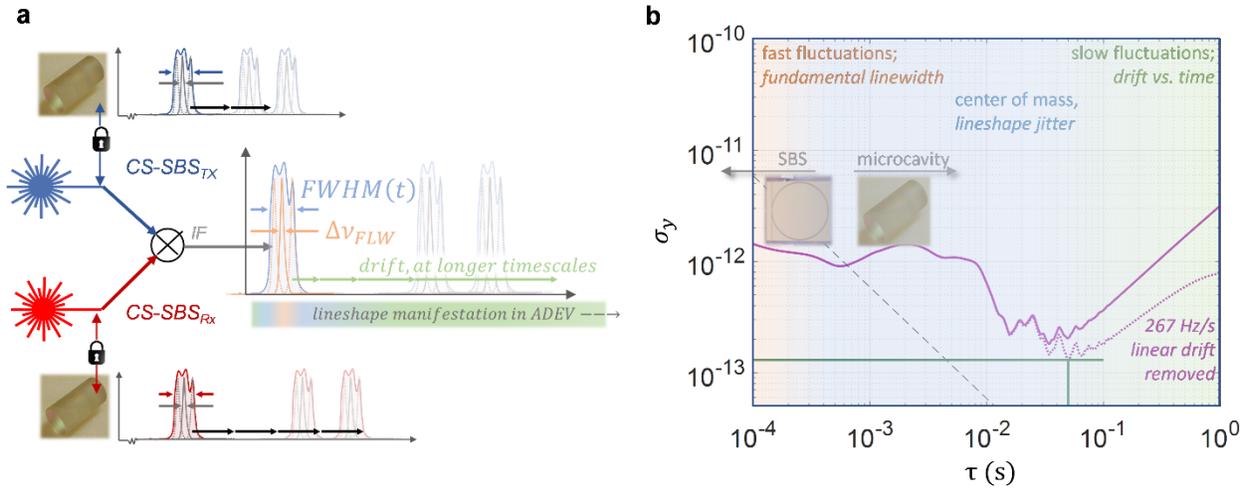

**Supplementary Fig. 2 | Heterodyne stability and Allan deviation. a**, Heterodyne beatnote at intermediate frequency (IF) from photomixing independent transmit (Tx) and receive (Rx) cavity-stabilized stimulated Brillouin scattering (CS-SBS) lasers using the electrical frequency discriminator (EFD). Lineshape as a function of observation time is illustrated, with fundamental linewidth ($\Delta\nu_{FLW}$) that broadens to an effective lineshape with full-width-half-maximum, $FWHM(t)$, as a function of observation time, $t$. **b,** Overlapping Allan deviation (ADEV) of the heterodyne beatnote with minimum fractional frequency instability (FFI) of < 2x10$^{-13}$ at 50 ms corresponding to <40 Hz. Raw ADEV FFI and linear drift removal (267 Hz/s) traces (solid, dotted respectively) are shown. The dashed (black) line represents the point where the RMS phase fluctuations are 1 rad for a given observation time, $\tau$ [7]

**Integral linewidth of heterodyne beatnote and optical carriers:** Using the EFD frequency noise, we calculate the integral linewidth of the optical heterodyne beatnote between Tx-Rx pump, Tx-Rx SBS, and Tx-Rx CS-SBS lasers using two methods, shown in Supplementary Fig. 3. The $\beta$-separation method[8] identifies regions of high modulation index that exceed the $\beta$-separation line[8] (gray dashed line in Supplementary Fig. 3a) and contribute to effective linewidth.

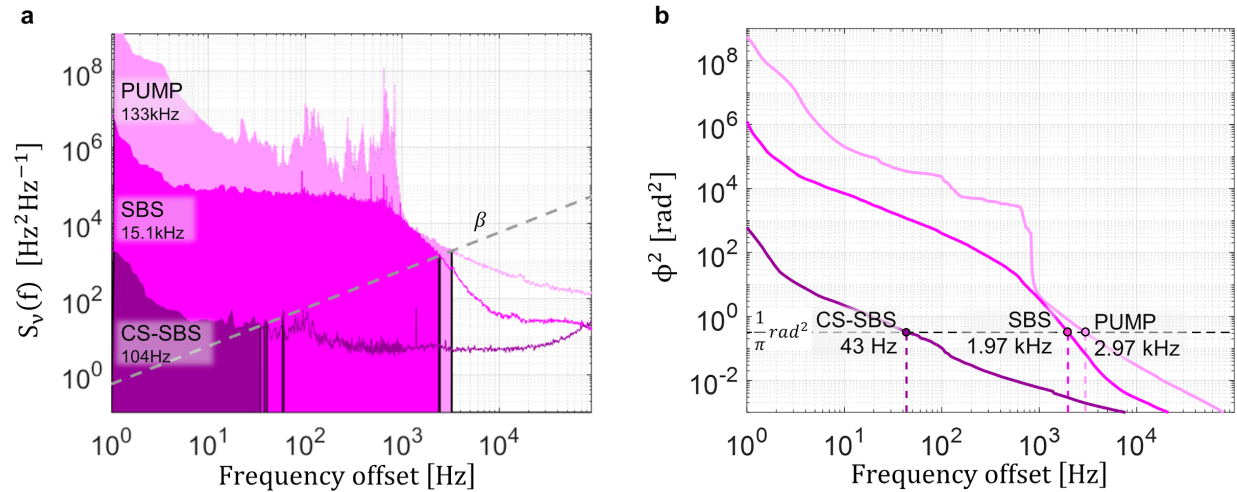

**Supplementary Fig. 3 | Integral linewidth. a**, Frequency noise spectra of the heterodyne beatnote between transmit and receive pump lasers (light purple), stimulated Brillouin scattering (SBS) lasers (medium purple), and cavity-stabilized SBS (CS-SBS) lasers (dark purple). Shaded regions correspond to high modulation index noise that exceeds the $\beta$-separation line and broadens the effective linewidth[8]. **b,** Integration of the frequency noise data in (a) from 10$^5$ Hz down to 1 Hz with the intersection of 1/π rad$^2$ corresponding to the effective linewidth[9,10].

ILWs are especially sensitive to high close-to-carrier noise and drift when using this method. We measure a reduction in the ILW of the pump laser heterodyne, 133 kHz ($\beta$-sep), down to 104 Hz ($\beta$-sep) for CS-SBS using this approach. The second method[9,10] is an integration of phase noise from high offset frequencies down to the offset frequency at which the integral equals $1/\pi$ rad$^2$, shown in Supplementary Fig. 3b. For a white-noise-only laser, this integration yields the offset frequency corresponding to the full-width-half-maximum (FWHM) of the laser lineshape. Using this method ($1/\pi$), we measure a reduction in ILW of the heterodyned pump lasers, 2.97 kHz, down to 43 Hz for the CS-SBS heterodyne. Assuming equal contribution of uncorrelated noise, we calculate independent CS-SBS optical ILWs ($1/\pi$) of ~30 Hz. Fundamental and integral linewidths are summarized in Supplementary Table 1.

# TABLES

**Supplementary Table 1 | Fundamental and integral linewidth summary.** Fundamental linewidths (FLW), as extracted by optical frequency discriminator (OFD) based frequency noise measurements, for transmit (Tx) and receive (Rx) pump, stimulated Brillouin scattering (SBS), and cavity stabilized SBS (CS-SBS) lasers. Integral linewidths as calculated using the integration of phase noise[9,10] and $\beta$-separation[8] methods using electrical frequency discriminator (EFD) derived frequency noise measurements on the Tx-Rx heterodyne beatnote for pump, SBS, and CS-SBS lasers.

|  | $\Delta\nu_{FLW}$ (OFD) | *ILW* (EFD, Tx-Rx beatnote) | |
|---|---|---|---|
|  | $\Delta\nu_{Tx}, \Delta\nu_{Rx}$ | ($1/\pi$) | ($\beta$-sep) |
| Pump | 90 Hz, 60 Hz | 2.97 kHz | 133 kHz |
| SBS | 1.4 Hz, 0.9 Hz | 1.97 kHz | 15.1 kHz |
| CS-SBS | 1.8 Hz, 1.8 Hz | 43 Hz | 104 Hz |